\documentclass[
preprint,
 amsmath,amssymb,
 aps,
]{revtex4-1}

\usepackage{graphicx}% Include figure files
\usepackage{dcolumn}% Align table columns on decimal point
\usepackage{bm}% bold math

\begin{document}

\title{Strong electron-lattice coupling as the mechanism behind charge
density wave transformations in transition-metal dichalcogenides}

\author{Lev P. Gor'kov} \email{gorkov@magnet.fsu.edu}
\affiliation{National High Magnetic Field Laboratory, Florida State
University, Tallahassee, FL 32310}

\begin{abstract}
We consider single band of conduction electrons interacting with
displacements of the transitional ions.  In the classical regime
strong enough coupling transforms the harmonic elastic energy for an
ion to the one of the well with two deep minima, so that the system is
described in terms of Ising spins.  Inter-site interactions order
spins at lower temperatures.  Extension to the quantum regime is
discussed.  Below the CDW-transition the energy spectrum of electrons
remains metallic because the structural vector $\mathbf{Q}$ and the FS
sizes are not related.  Large values of the CDW gap seen in the
tunneling experiments correspond to the energy of the minima in the
electron-ion two-well complex.  The gap is defined through the density
of states (DOS) inside the electronic bands below the CDW transition.
We focus mainly on electronic properties of transition-metal
dichalcogenides.
\end{abstract}

\maketitle

\section{Introduction}

Origin of charge density waves (CDW) in the transition-metals
dichalcogenides (TMDC) is subject of debates since their discovery in
the early 70's \cite{cit1}.  At the time, the most popular theoretical
scenario was the so-called ``nesting'', the congruency of two or more
Fermi surfaces (FS) separated by a vector $\mathbf{Q}$ in the momentum
space.  Interactions with the momentum transfer $\mathbf{Q}$ would
lead to the CDW instability (with $\mathbf{Q}$ becoming the structural
vector) and to the opening of energy gaps on FS.

The alternative explanation \cite{cit2} ascribes the structural
instability to the presence of saddle points in the electronic
spectrum near the Fermi energy.  The logarithmic singularities in the
electronic density of states (DOS) at the saddle points favor
instabilities with the momentum transfer, $\mathbf{Q}$ connecting the
two saddle points.  The mechanism also leads to the energy gaps in the
electron spectrum.

Subsequent experiments and band structure calculations for real
materials gave no support to either of the two concepts.  The CDW
transition does not affect properties of dichalcogenides any
noticeable.  Materials remain metallic below the transition
temperature, $T_\mathrm{CDW}$.  Superconductivity in 2$H$-TaSe$_2$,
2$H$-TaS$_2$ and 2$H$-NbSe$_2$ takes place on the background of the
CDW phase.  The CDW gap values $\sim0.1$~eV surprisingly large in
comparison with the values of transition temperatures,
$T_\mathrm{CDW}(\mathrm{Ta}) = 122$~K and $T_\mathrm{CDW}(\mathrm{Nb})
= 33.5$~K, were observed via the $dI/dV$ characteristics in the
tunneling experiments \cite{cit5}.  The gap of such order of magnitude
is noticeable in the infrared data \cite{cit6} (2$H$-TaSe$_2$) even
close to the transition temperature.

The phenomenological analysis in \cite{cit7} revealed the short
coherence length in the CDW phase, ${\xi_0} \sim 3 - 10$~\AA, thus
implying the important role of fluctuations.

The density functional calculations \cite{cit3} (2$H$-NbSe$_2$) have
shown that the contribution from the nested pieces of FS into the
charge susceptibility is negligible at the expected value of the
vector $\mathbf{Q}$, although, indeed, the ground state of the system
at $T = 0$, according to \cite{cit3}, is the CDW phase.

It was well known since 1958 that at the strong enough \emph{e-ph}
coupling the square of renormalized phonon frequency turns out
negative, thus signaling instability of the lattice \cite{cit18}.  In
\cite{cit4} through the example of 1$T$-TaS$_2$, it was demonstrated
that CDW transitions in the transition-metal dichalcogenides can be
driven by the Migdal's instability \cite{cit18}.  The new tinge added
to this concept in \cite{cit4} was that dispersion of the renormalized
(imaginary) soft mode can become strongly peaked near the structural
vector of instability, $\mathbf{Q}$ if the \emph{e-ph} matrix element
depends on the transferred momentum.

There is no small physical parameter in the problem \cite{cit18}.
Correspondingly, there are no other means in \cite{cit4} to discuss
electronic properties of 2$H$-TaSe$_2$ and 2$H$-NbSe$_2$ beside
numerical calculations.

As it was already mentioned above, the nesting scenario and that one
of Ref.  \onlinecite{cit2} both contradict to the experimental data
\cite{cit5,cit6} and to their analysis and the interpretation
\cite{cit4,cit7}.  In spite of that, the two concepts are commonly
used for interpretation of results even in recent ARPES experiments
(see, e.g., \cite{cit19,cit24}).

We address peculiarities of transition-metal dichalcogenides from the
point of view differing from \cite{cit4}.  We argue that instead of
development of the Migdal instability in the soft mode scenario
\cite{cit4} for the \emph{propagating} phonon mode, the CDW
instability may realize itself in the two stages.  Namely, at first,
strong \emph{e-ph} interactions bring electrons and ions close
together in kind of a polaronic effect similar to the one first
discussed for V$_3$Si in \cite{cit9}.  The potential in which the ion
moves becomes anharmonic.  At a strong enough \emph{e-ph} coupling the
potential possesses two minima, so that the system can be described in
terms of Ising spins.  The subsequent ordering of the local sites at a
lower temperature occurs due to inter-site interactions.

In Ref.  \onlinecite{cit9} electron-lattice interactions were limited
to the interaction of electrons with single dispersionless (Einstein)
optical mode.  In the approach below conduction electrons interact
with arbitrary ionic displacements.  Strength of the intersite
interactions and the value of contributions into the local elastic
potential are separately evaluated.

``Trapping'' of the electronic cloud near an ion by the elastic field
is deemed responsible for the main energy gains and pre-determines,
whereby, by the order of magnitude, the energy scale of the CDW gap.
According to our estimates, the wells' minima are deep ($\sim$ few
tenths of 1~eV).  Thereby, the Kondo-like regime of quantum tunneling
that was the subject of the main concern in \cite{cit9} seems to be of
no relevance to dichalcogenides, at least at the temperatures of
interest.  We focus in the following on properties of the electronic
sub-system.

\section{Choice of the model}

Before proceeding further, we briefly summarize the information
pertaining to the CDW transitions in TMDC.

Symmetry of the order parameters governing the transition was
discussed in \cite{cit7,cit8}.  The lattice superstructure below
$T_\mathrm{CDW}$ \cite{cit1} is formed by the triple-$\mathbf{Q}$
modulations of the ionic positions along the three symmetry axes in
the hexagonal \textit{2H}-phase.  The $\mathbf{Q}$-vector may be
incommensurate, but even then $\mathbf{Q}$ remains very close to
$\vec{a}^{\ast}/3$ (here $a^{\ast}=4\pi /\sqrt 3 a)$.

It was perceived already in the early 70's \cite{cit10} that the
structural changes in 2$H$-NbSe$_2$ are related to displacements of
the niobium ions.  Correspondingly, in what follows, we choose a
simplified model and consider only shifts of the transition-metals
ions (Nb or Ta) along each of these three symmetry lines.  The
dramatic softening of the longitudinal acoustic mode with $\mathbf{Q}$
about $\vec{a}^{\ast}/3$ was recently observed in \cite{cit11}.  Note,
in passing, that the optical modes and the longitudinal acoustic
branch of the same $\Sigma_1$- symmetry are actually linearly coupled
near point $\vec{a}^{\ast}/3$ of the Brillouin Zone (BZ).  We return
to this later to show that softening of the optical mode is
accompanied by softening of the acoustic phonon, and vice versa.

Transition-metals dichalcogenides are layered compounds with the weak
Van der Waals coupling between the layers.  We restrict ourselves by
the quasi-two-dimensional properties (Q2D) of a single layer.

We consider a single isotropic band of electrons interacting strongly
with displacements of the transition-metal ions.  Implicitly, the band
is assumed to be predominantly of the 4f (Nb) - or 5f (Ta) - character.

In the model \cite{cit4} the ``bare'' frequency of the
\emph{propagating phonon mode}, $\omega_0^2 (\mathbf{k})$ is
renormalized by the polarization operator \cite{cit12} (with the
\textit{e-ph} matrix element $g(\mathbf{p},\mathbf{p+k})$ is peaked at
$\mathbf{k} \approx \mathbf{Q}$).  Unlike \cite{cit4}, our attention
is concentrated on the \emph{local environment} of the single ion.  In
this respect, as we discuss later, our model shows some features
common to the Holstein model \cite{cit13}.

For a heavy ion one may expect that quantum effects are not crucial
for properties of the lattice, at least, at not too low a temperature.
The temperature of the (incommensurate) transition in 2$H$-TaSe$_2$,
$T_\mathrm{CDW} = 122$~K \cite{cit1} is rather high and, in the first
approximation, one can neglect the kinetic energy of the Ta ions.
 
With the notations $u_i$ for a displacement of the single ion at the
point $R_i$ the ``bare'' elastic matrix is: $U(R_i - R_j) u_i u_j$.
At $i=j$ it is merely the potential energy of the oscillator:
\begin{equation}
\label{eq1}
U(R_i - R_j ) u_i u_j \Rightarrow \frac{1}{2} M \omega_0^2 u_i^2 
= \frac{1}{2} k u_i^2~.
\end{equation}
Interactions with electrons change the elastic matrix.  We calculate
the contribution to the elastic energy that is due to interactions of
electrons with arbitrary static displacements.  

With the Hamiltonian for the \emph{e-ph} interaction in the form
\begin{equation}
\label{eq2}
\hat{H}_{e-ph} = \sum_i g u_i  \hat{\psi}^{+} (R_i) \hat{\psi}(R_i)
\end{equation}
the energy of the coupled electron-lattice system $E(u_1 ...  u_N;g)$
can be calculated from the equation:
\begin{equation}
\label{eq3}
\frac{\partial E(u_{1} ...u_{N} )}{\partial g}=\sum_i u_i  n(R_i)~,
\end{equation}
where $n(R_i)= \langle \hat{\psi}^{+}(R_i) \hat{\psi}(R_i) \rangle$ is
the number of electrons per unit cell at the point $R_i$.  In terms of
the electronic Green function (at $T=0$, \cite{cit12}), $G_{\alpha
\beta }(R,t;R',t') = -i \langle T(\psi _{\alpha} (R_i,t) \hat{\psi
}_{\beta }^{+}(R',t') )\rangle$, one has: $n(R_i) = - 2 i
G(R,t;R,t+\delta)$.  (The Green function is diagonal in the spin
indices: $G_{\alpha \beta }(R,t;R',t') = -i \delta_{\alpha \beta }
G(R,t;R',t')$ ).

\section{Coupled electron-lattice system in the classical regime}

At static displacements, one needs to know only the frequency
component, $G(R,R';\omega)$:
\begin{equation}
\label{eq4}
G(R,t;R',t') = \int \frac{d \omega}{2 \pi}\, G(R,R';\omega)
\exp(-i \omega (t-t'))~.
\end{equation}
Re-arrange the power expansion of $G(R,R';\omega)$ in $u_i$ as: 
\begin{widetext} 
\begin{eqnarray}
    \label{eq5}
    G(R,R';\omega )&=& G_0 (R-R';\omega )+\sum_i G_0(R-R_i;\omega) g
    \bar{u}_i G_0 (R-R_i;\omega) \\
    &+& \sum_{i \ne k} G_0(R-R_i;\omega) g \bar{u}_i
    G_0(R_i-R_k;\omega) g \bar{u}_k G_0(R_k - R';\omega)+...~.
    \nonumber
\end{eqnarray} 
\end{widetext}
The \emph{e-ph} contributions in \emph{all powers} in $g$ are now
summed first for the single site: $u_i \Rightarrow \bar{u}_i$.  Each
of $\bar{u}_i, \bar{u}_k , ...  $ in Eq.\ (\ref{eq5}) stands for the
``dressed'' local deformation at the corresponding site, $R_i$, $R_k$,
..., and is determined by the relation:
\begin{equation}
\label{eq6}
\    bar{u}_i = u_i + u_i G_0(R_i = R_i;\omega) g\bar{u}_i~. 
\end{equation}
The free Green function, $G_0 (R_i=R_i;\omega)$ equals \cite{cit12}:
\begin{widetext}
\begin{equation}
\label{eq7}
    G_0 (R_i =R_i;\omega)=\int \frac{d^2 \bar{p}}{(2\pi)^2}\, 
    \frac{1}{\omega - E(p)+ E_F +i \mathrm{sign}(\omega) \delta}~.
\end{equation}
\end{widetext}
With the help of identity 
%%
%%
%%\begin{widetext}
    $$\frac{1}{\omega - \xi + i \mathrm{\mathrm{sign}}(\omega) \delta}= 
    P\left(\frac{1}{\omega -\xi}\,\right) 
    -i \pi \mathrm{\mathrm{sign}}(\omega) \delta(\omega -\xi )$$
%%\end{widetext}
%%
%%
and assuming the electron-hole symmetry for $|\xi| = |E(p)-E_F| $, one
obtains:
\begin{equation}
\label{eq8}
    G_0 (R_i =R_i;\omega) = -i \pi \nu (E_F)
    \mathrm{\mathrm{sign}}(\omega)~.
\end{equation}
After trivial calculations, one finds the electronic contribution into
the local elastic matrix at the site $R_i$, $E(u_i) = -2 i \int_0^1 dg
u_i G(R_i =R_i; t=t'+\delta ;g)$:
\begin{equation}
\label{eq9}
    E(u_i) = - \frac{W}{2\pi}\ln\left\{1+\left[\pi \nu (E_F) g
    u_i\right]^2\right\}~,
\end{equation}
where $W$ is the bandwidth.  $E(u_i)$ is determined as the energy per
one ion, $\nu (E_F)$ is the number of states at the Fermi level per
unit cell: $\nu (E_F)=\left(\frac{m}{2\pi }\,\right) S_0$, where $S_0$
is the area of the 2D unit cell.  $W \nu (E_F)=s$ is an insignificant
model parameter, $s \sim 1$.  We take $s=2$.

The interaction between two sites, $E(u_i ,u_k)$, is given by the
second term in Eq.\ (\ref{eq5}):
\begin{equation}
\label{eq10}
    E(u_i ,u_k)=u_i u_k (-i)g^2 \int \frac{d \omega}{2\pi }~, G_0^2
    (R_{i,k} ;\omega )
\end{equation} 
where $R_{i,k} \equiv |R_i -R_k|$.

The analytic form of $G_0 (R_{i,k};\omega)$ being cumbersome in 2D,
for the estimate we use its asymptotic at $p_F R_{i,k} > 1$:
\begin{widetext}
\begin{equation}
\label{eq11}
    G_0(R;\omega ) \Rightarrow i \nu(E_F) \sqrt{\frac{2\pi }{p_F R}\,}\\
    \exp\left\{i \mathrm{sign}(\omega) \left[\left(p_F + \frac{\omega}
    {v_F}\right)R - \frac{\pi}{4}\right]\right\}~.
\end{equation}
\end{widetext}
We obtain:
\begin{equation}
\label{eq12}
    E(u_i,u_k)=u_i u_k g^2 \nu^2(E_F) p_F v_F \frac{\sin(2 p_F
    R)}{(p_F R)^2}~.
\end{equation}
The total elastic energy of a single ion is the sum of Eqs. 
(\ref{eq1}) and (\ref{eq9}).  Introducing:
\begin{equation}
\label{eq13}
    g^2=\left(\frac{M\omega_0^2}{2 \pi \nu (E_F)}\, \right) 
    \Lambda^2~,
\end{equation}
\begin{equation}
\label{eq14}
    \frac{\pi }{2}\, \left(M \omega_0^2 \nu (E_F)\right) = 
    \frac{1}{u_0^2}~,
\end{equation}
and the dimensionless notations for the ions' shifts,
$\tilde{u}_i=(u_i/u_0)$, one writes down the local elastic energy in
the following simple form:
\begin{equation}
    \label{eq15}
    U_\mathrm{tot} (u_i) = \frac{1}{\pi \nu (E_F)}
    \,\left[\tilde{u}_i^2 -\ln \left(1+\Lambda^2\tilde{u}_i^2
    \right)\right]
\end{equation}
Here $\Lambda^2$ is the square of the dimensionless \textit{e-ph}
coupling constant and $u_0$ determines the spatial scale of the local
elastic potential.  The energy scale $T^{\ast }=1/\pi \nu (E_F) = W/2
\pi$ is expressed through DOS in the electronic band.  For
2$H$-NbSe$_2$ the band calculations \cite{cit14} gave $\nu (E_{F}) =
2.8$~states/eV per two bands.  Here $T^{\ast }$ is of the order of
tenths of 1~eV.
 
At $\Lambda^2 > 1$ the potential $U_\mathrm{tot} (u_i)$ has two deep
minima at $\tilde{u}_{+,-} = \pm \sqrt {1-\Lambda^{-2}}$:
\begin{equation}
    \label{eq16}
    U_\mathrm{tot} (u_{+,-} )=(1/\pi \nu (E_F))
    \left[1-\Lambda^{-2}-\ln \Lambda^2\right]~.
\end{equation}

At $\Lambda^2 < 1$ and temperatures below $T^{\ast}$ $\tilde{u}_i \ll
1$ and $U_\mathrm{tot}(u_i)$ can be written as:
\begin{equation}
    \label{eq17}
    U_\mathrm{tot}(u_i)=(1/\pi \nu (E_F)) \left\{\tilde{u}_i^2
    (1-\Lambda^2) + \frac{\Lambda^4}{2}\, \tilde{u}_i^4 \right\}~.
\end{equation}
The quartic term in Eq.\ (\ref{eq17}) is small, but the anharmonic
contribution into the elastic energy is the necessary ingredient in
the molecular field approach to a CDW transition (see, e.g.,
\cite{cit15}).

Re-writing $p_F v_F$ as: $p_F v_F = (1/\pi \nu (E_F))(p_F^2S_0/2)$,
where $S_0 = (\sqrt{3}/4) a^2$ is the area of the triangular unit
cell, Eq.\ (\ref{eq12}) can be written down in the notations of Eqs.\
(\ref{eq13},\ref{eq14}):
\begin{widetext}
\begin{equation}
    \label{eq18}
    E(u_i ,u_k )=\frac{1}{\pi \nu (E_F )}\, \left[ \frac{\sqrt{3}
    }{4\pi^2}\, \Lambda^{2} \tilde{u}_i \tilde{u}_k
    \left(\frac{a}{R}\,\right)^2 \sin (2p_F R) \right]~.
\end{equation}
\end{widetext}
The numerical factor in this expression shows that the inter-site
interactions are weak compared to the on-site $U_\mathrm{tot} (u_i)$.

At $\Lambda^2 > 1$ the model reduces to the model of interacting Ising
spins:
$$\tilde{u}_{+,-} (i)\Rightarrow \sqrt{1-\Lambda^{-2}}\, \sigma_{i}, \ \ \sigma_{i} = \pm 1~.$$
 
When $\Lambda^{2} < 1$, the expressions for $U_\mathrm{tot} (u_{i} )$
and Eq.\ (\ref{eq18}) for $E(u_{i} ,u_{k})$ together complete
formulation of the problem: in the classical regime the system is
fully described by the partition function $Z(T,g)$:
\begin{equation}
    \label{eq19}
    Z(T,g) = \int (\Pi du_i) \, \exp \left[-\frac{1}{T}\, \sum_{i,k}
    U_\mathrm{tot}(u_i, u_k)\right]~,
\end{equation}
where $U_\mathrm{tot} (u_i, u_k) = U_\mathrm{tot} (u_i) + E(u_i, u_k)$. 

The phase transitions are habitually treated in the molecular field
approximation (see, e.g., \cite{cit15}).  The method being not exact
for short-ranged interaction; we do not discuss details of the CDW
transitions itself.

For local properties one has for the partition function, $Z(T,g)$:
\begin{widetext}
\begin{equation}
    \label{eq20}
    Z_{i} (T,g) = \int {d\tilde{u}_i} \exp \left\{-\frac{1}{\pi \nu
    (E_F)T}\, \left[\tilde{u}_i^2 - \ln (1+\Lambda^2
    \tilde{u}_i^2)\right]\right\}~.
\end{equation} 
\end{widetext}
Scattering of electrons on the lattice displacements above
$T_\mathrm{CDW}$ is characterized by the average $\langle g \bar{u}_i
\rangle$ that enters the denominator of the Green function:
\begin{equation}
    \label{eq21}
    G^{-1}(R;\omega) = \omega - E(p) + E_F + \langle g \bar{u}_i
    \rangle~.
\end{equation}
From Eq.\ (\ref{eq6}) it follows: $g \bar{u}_i = gu_i [ 1-i
\mathrm{sign}(\omega) \pi \nu (E_F) g u_i ]^{-1}$.  In the normal
phase terms that are odd in $u_i$ can be omitted.  In dimensionless
variables:
\begin{equation}
    \label{eq22}
    g \bar{u}_i = i \mathrm{sign}(\omega)\, \frac{1}{\pi \nu
    (E_F)}\,  \frac{\Lambda^2\tilde{u}_i^2}{1+\Lambda^2
    \tilde{u}_i^2}~.
\end{equation}
At $\Lambda^2 < 1$ the imaginary part is: $G^{-1}(R;\omega )$ is
$\langle g\bar{{u}}_{i} \rangle \cong i \mathrm{sign}(\omega)
\Lambda^2(\pi /2)\nu (E_F )T$.  In the classical regime above
$T_\mathrm{CDW}$ the resistivity of the system would be linear in $T$.
In the opposite limit of $\Lambda^2 > 1$, $\tilde{u}^2 =
\tilde{u}_{+,-}^2 = 1 - \Lambda^{-2}$, and the imaginary part is a
constant of order of $1/\pi \nu (E_{F})$.  The entropy for the system
of non-interacting Ising spins is finite.
 
In the ordered state with all ions occupying same minima, non-zero
$g\bar{u}_i \ne 0$ stands together with the chemical potential:
\begin{equation} 
\label{eq23} 
    g\bar{u}_i = \pm \frac{1}{\pi \nu (E_F)} \sqrt{\Lambda^2-1}~.
\end{equation}

So far, for simplicity of the arguments, it was tacitly implied that
the order parameter $g\bar{u}_i \ne 0$ in Eq.\ (\ref{eq22}) stands for
the CDW transition with the structural vector $\mathbf{Q}=0$.  As in
TMDC the $\mathbf{Q}$-vector is non-zero, the order parameter
$g\bar{u}_i(Q)$ in Eq.\ (\ref{eq21}) couples electronic states with
the energies $E(p)$ and $E(p+Q)$.

In such a way, non-zero $g\bar{u}_i(Q)$ changes the energy spectrum
and, hence, DOS in the vicinity of such paired points, but does it
mainly for energies away from the Fermi level because in
dichalcogenides the $\mathbf{Q}$-vector and sizes of the
Fermi-surfaces are not related.  Of course some Fermi-surfaces’
points may be affected at onset of the CDW order, as it is really
observed in \cite{cit19}.

\section{Discussion of the results}

\subsection{CDW instability: above and below $T_{CDW}$}

It is interesting to discuss the relevance of the above model to the
experimental results, in particular, for 2$H$-TaSe$_2$.

First, recall that the non-linear potential (\ref{eq9}) was derived
for a single ion.  For Eq.\ (\ref{eq9}) to be meaningful, $u_0$ must
be small compared to the lattice parameter, $a \approx 3.45$~\AA. With
the phonon frequencies, $\omega_0$ for NbSe$_2$ typically in the range
$10-20$~meV, is indeed small: ${u_0} \leqslant 0.30 - 0.15$~\AA.
Similar estimates for 2$H$-TaSe$_2$ give ${u_0} \leqslant 0.2 - 0.1$~\AA. 
This justifies the assumption.  Experimentally, the
superlattices’ shifts of cations are between 0.1~\AA\ \cite{cit1} and
0.5~\AA\ \cite{cit9} and, hence, have same order of magnitude as
${u_0}$, as it should be in the model of Ising spins.

In \cite{cit6} the large CDW gaps $\sim 90$\ meV and $\sim 60$\ meV
for 2$H$-TaSe$_2$ and 2$H$-NbSe$_2$, respectively, were derived from
the tunneling non-linear current characteristics, $dI/dV$.  Such
values are in the right correspondence with the energy scale, ${T^*} =
1/\pi \nu ({E_F})$ equal to $0.1- 0.3$~eV, depending on the number of
states per unit cell in the specific material (for NbSe$_2$ $\sim ~
2.8$~eV$^{-1}$ per unit cell per two Nb bands \cite{cit14}).  So far,
however, the CDW gap itself was rather vaguely defined.  There, it was shown
that the CDW with non-zero $\mathbf{Q}$ couples the electronic states
with $E(p)$ and $E(p+Q)$.  For the two symmetric points, $E(p-Q/2)$
and $E(p+Q/2)$, say, inside the energy band for the Fermi surface
centered at the $\Gamma$-point, the energy spectrum in their vicinity
is:
\begin{equation} 
\label{eq23a} 
{\bar E_{1,2}}(p \pm Q/2) = E(p \pm Q/2) \pm |g{\bar u_i}(Q)|~.
\end{equation}

Eq.\ (\ref{eq23a}) defines the CDW gap as it appears in the $I-V$
characteristics which measure the energy dependence of DOS. The CDW
gaps depend on the value of the parameter, ${\Lambda ^2}>1$ as in
Eqs.\ (\ref{eq15}) and (\ref{eq22}).

In 2$H$-NbSe$_2$ the vector ${\mathbf{Q}} \approx {\vec{a}^*}/3$ is
shorter than the radius of the Fermi surface centered at the
$\Gamma$-point.  Changes in DOS at the CDW transition for the three
symmetric pairs of points \emph{inside} the Fermi surface were
directly detected in the ARPES experiment \cite{cit25}.  (Actually, in
\cite{cit25} the map of all the spots paired with the Q-vector was
obtained for the whole BZ; the pattern has the hexagonal symmetry).

Recall now that theoretically the frequency of the collective
excitations behaves differently at $T=T_\mathrm{CDW}$ for the local
potentials with one or two minima \cite{cit16} (see also in
\cite{cit15}).  In the former case the frequency vanishes at the
temperature of the transition.  In the latter, the frequency remains
finite as an ion is now ``trapped'' by one of the two minimum.
While neutron experiments \cite{cit1}(b) for
2$H$-TaSe$_2$ gave finite $\omega^2(Q) \approx 20$~meV$^2$ at
$T_\mathrm{CDW} = 122$~K, softening of the acoustic phonons was
observed for 2$H$-NbSe$_2$ at $T_\mathrm{CDW}=33.5$~K \cite{cit11}.
 
The difference in the phonon modes' behavior in the two materials
needs a clarification.  Recall that the longitudinal acoustic branch
is coupled linearly with the optical mode, $u$ of the same $\Sigma_1$-
symmetry at $\mathbf{Q} =\vec{a}^{\ast}/3$.  Free energy then has a
contribution of the form: $F(u,s)=\omega_s^2 (s^2/2)+\omega_u^2
(u^2/2)+ t u s$, where $s$ stands for the acoustic branch.  Minimizing
$F(u,s)$ and excluding $u$ gives $F(u,s)=[\omega_s^2
-(t^2/\omega_u^2)](s^2/2)$.  Let the optical mode, $u$ be the mode
that drives the transition.  At $\omega_u^2 (T)\to 0$ (one minimum),
the acoustic mode is the first one that manifests the onset of the
transition.  If the potential has a few minima, the effective
frequency of the acoustic mode $\omega_\mathrm{s,eff}^2 =[\omega_s^{2}
-(t^2/\omega_u^2)]$ at $T_\mathrm{CDW}$ may or may not be zero
depending on the temperature behavior of $\omega_u^2(T)$.  Data
\cite{cit1}(b) for 2$H$-TaSe$_2$ agree better with the second
possibility.

The two 2$H-$materials have different masses of Ta- and Nb- ions.  It
is known \cite{cit16} that at lower temperatures, when quantum effects
prevail, $\omega_u^2 (T)\to 0$ even for a potential with a few minima.
This could be another possible interpretation for the acoustic phonon
frequency vanishing at $T_\mathrm{CDW} = 33.5$~K in the 2$H$-NbSe$_2$
\cite{cit11} and the finite $\omega^2(Q)$ at $T_\mathrm{CDW} = 122$~K
in 2$H$-TaSe$_2$ \cite{cit1}(b).

The thermodynamics of the second order phase transition would look
much alike for the displacive transition \cite{cit4} or for the
ordering of the Ising spins.  The finite ${\omega ^2}(Q)$ at
$T_\mathrm{CDW} = 122$~K in 2$H$-TaSe$_2$ \cite{cit1}(b) is the first
argument supporting the Ising spins model.

We argue that the temperature dependence of resistivity in all TMDC's
seems to indicate in the same direction.  In fact, for a broad
temperature interval the resistivity, $\rho(T)$ behaves as $\rho (T) =
{\rho _0} + aT$, with the large intercept ${\rho _0} \sim 100 -
150$~$\mu\Omega\cdot\textrm{cm}$ (2$H$-TaSe$_2$, \cite{cit22}(a, b)).
This behavior is consistent with the inverse mean free time in Eq.\
(\ref{eq21}) for the Green function of the form ${\tau ^{ - 1}} = \tau
{}_0^{ - 1} + \bar aT$ and $\tau _0^{ - 1} = (1/\pi \nu ({E_F}))(1 -
{\Lambda ^{ - 2}})$.

In the ARPES experiment \cite{cit24} the self-energy, 
$\Sigma (\omega ) = \Sigma '(\omega ) + i\Sigma''(\omega )$ 
was measured directly on the Fermi surface of 2$H$-TaSe$_2$ centered at the
$\Gamma$-point of BZ.  The imaginary part, $\Sigma ''(\omega ) \approx 
60-70$~meV is frequency independent above $T_\mathrm{CDW}$; this value is 
consistent with $\tau_0^{ - 1} \sim {T^*} = 1/\pi \nu ({E_F})$.  An estimate 
with $\tau _0^{ - 1} \sim {T^*}$ gives the right value of $\rho _0$ from \cite{cit22}(a,b).

As to the real part, $\Sigma '(\omega )$ is negligible at $T=111$~K
(i.e. above $T_\mathrm{CDW} =88$~K).  Below $T_\mathrm{CDW}$ $\Sigma
'(\omega )$ starts to show the behavior typical for the
self-energy of normal electrons interacting with phonons \cite{cit18}
and is peaked at the phonons’ frequencies $\sim 50$~meV. Such
typical metallic signature obviously agrees with the suggestion that
the CDW order parameter $g\bar{u}_i(Q)$ in Eq.\ (\ref{eq23}) affects
the electronic states of the energy band at the $\Gamma$-point only
below the Fermi level \cite{cit23,cit26}.

Finally, the NMR methods may help with revealing properties of the
transition atoms occupying the symmetry positions in the commensurate
phase of 2$H$-TaSe$_2$, as it is pointed out below.
 
\subsection{Quantum regime}

At low $T$ quantum effects become important in few aspects.  Consider
first the Schr\"{o}dinger equation for an ion moving in the rigid
potential $U_\mathrm{tot}(u_i)=(1/\pi \nu (E_F))
\bar{U}(\tilde{u}_i)$, with $\bar{U}(\tilde{u}_i)$ in the
dimensionless notations.  With the kinetic energy in the same
notations, one has:
\begin{equation}
    \label{eq24}
    -\frac{1}{2\bar{M}}\, \frac{d^2}{d \tilde{u}^2}\, \Psi(\tilde{u})+
    [\bar{U}(\tilde{u})-\bar{E}]\Psi(\tilde{u})=0
\end{equation} 
The dimensionless ``mass'', $\bar{M}$ in (\ref{eq24}) is defined 
by $1/\bar{M}=[\pi \nu (E_F )\omega_0]^2/2$. (The adiabatic 
parameter $1/\sqrt{\bar{M}} =(\pi \nu (E_F )\omega_0 )/\sqrt{2} $ is 
about one tenth in 2$H$-NbSe$_2$). 

For the single-minimum well ($\Lambda^2 < 1$) quantum effects are
important when $T_{CDW} < \omega_0$.  The lattice oscillations are now
quantized and the electronic Green function is ``dressed'' by phonons.
In the adiabatic approximations, solution for the problem of
interacting electrons and phonons in normal metals was given many
years ago in \cite{cit18}.

At first glance, Eq.\ (\ref{eq24}) adds a possibility of quantum
tunneling between minima of the two-well potential \cite{cit16}.
Actually, Eq.\ (\ref{eq24}) could account only for tunneling in a
``rigid'' potential, a potential built in the lattice only by the
inter-atomic forces.  The two-well potential (Eqs.\
(\ref{eq9},\ref{eq15})) is formed by interactions between electrons
and the local lattice distortions.  It was emphasized in \cite{cit9}
in connection with the martensitic transitions in Nb$_3$Sn and
V$_3$Si, that at the tunneling event, at which the ion goes over, say,
from ${u_ + }$ to ${u_ - }$, the electronic configuration reverses as
well.  Such a feature cannot be described in terms of Eq.\
(\ref{eq24}).  According to \cite{cit9}, the under-barrier tunneling
results in a Kondo-like quantum regime at which the height of the
barrier would diminish with temperature.  The parameter $\sqrt {\bar
M} $ stands in the exponent of the expression for the tunneling matrix
element.  With no closed solution found in \cite{cit9} one may only
argue that with the value of $\sqrt {\bar M} \sim 10$ the Kondo-like
regime \cite{cit9} is probably of no relevance to the CDW physics in
TMDC.

Below the transition in the CDW ground state ions will occupy their
proper minima, in accordance with the superlattice pattern
\cite{cit1}(b) and tunneling between minima must stop.  Nevertheless,
experimentally, the pattern reveals an interesting peculiarity.
Indeed, in the CDW phase one in three atoms along the symmetry lines
finds itself in the position with the trigonal symmetry (2$H$-TaSe$_2$
\cite{cit1}(b); see Fig.\ 4 in \cite{cit8}).  While below transition
the intersite interactions do indeed arrest quantum tunneling between
minima for the other two of the three atoms, the degeneracy is not
lifted for the atom in this symmetric position.

NMR experiments seem be able to verify the very concept of the two
minima-potential by studying the ions in positions with the trigonal
symmetry.
 
According to ARPES data \cite{cit19}, the CDW transition in
2$H$-NbSe$_2$ slightly affects only the two-barrel Fermi surface at
the K-point in BZ, while FS at the $\Gamma $-point remains intact.
The three $Q$-vectors couple together the three points on the inner FS
at K, which are seen experimentally with small but observable gaps at
these points.  Surprisingly, the observation of such small local gaps
($\sim 2.4$~meV) was interpreted in \cite{cit19} as the conclusive
proof in favor of the nesting mechanism of the CDW formation.

The physics of strong local \emph{e-ph} interaction bring us back to
the Holstein model \cite{cit13} of electrons interacting with
dispersionless phonons.  No exact solution is known for the Holstein
model either, but its low temperature physics was investigated
numerically in DMFT (Dynamical Mean Field Theory, \cite{cit20})
approximation.  (For a brief summary of results for the Holstein model
at $T=0$ see \cite{cit21}).

The DMFT approach is strictly local.  Intersite interactions (see Eq.\
(\ref{eq18})) and the CDW transition itself cannot be treated by DMFT.
Here we indicate parallels between our physics above and the results
\cite{cit21} at $T=0$ for the Holstein model.  Among them are: (1) The
double-well potential that develops when the \emph{e-ph} coupling
constant becomes larger some critical value; (2) Large imaginary part
in the Green function: Spectral Function extends over the energy
interval that significantly exceeds the phonon frequency (compare with
our Eq.\ (\ref{eq22})); (3) The ground state remains metallic (at
least for not-too-strong \emph{e-ph} coupling).

In \cite{cit21} the mass of electronic excitations increases as the
residue at the pole of the electronic Green function decreases.

Judging by these results, the physics studied above allows its
extension into the low temperature regime.

\section{Summary}

In summary, we calculated the local elastic energy for single ion by
re-summing exactly the electron-lattice interactions in the real
space.  Strong \emph{e-ph} interactions qualitatively change the local
environment by binding ions and electrons together, thus breaking the
adiabatic approximation.  The concept is the realization of the Migdal
instability different from that one considered in \cite{cit4}.  The
CDW transition takes the form of a phase transition in the system of
interacting Ising spins.  The value of the structural vector
$\mathbf{Q}$ and parameters of the Fermi surfaces being not related
with each other, the energy spectrum of electrons remains metallic
below the temperature of the CDW transition.  The CDW gap seen in the
tunneling experiments below the transition is defined in terms of
reduced DOS inside the electronic bands.  Its large value is
consistent with the energies of the deep minima of the local
potential.  Other experimental evidences in favor of the Ising model
were enumerated.  The electronic properties of the transition-metal
dichalcogenides, as they are seen by ARPES, agree well with the 
suggested concept.

Conclusions from the analytical results derived in the classical
regime can be extended to lower temperatures as it follows from the
comparison with numerical results for the Holstein model.

\section{Acknowledgments}

The author thanks T. Egami and V. Dobrosavljevic for helpful
discussions.  The work was supported by the NHMFL through NSF Grant
No.  DMR-0654118 and the State of Florida.


\begin{thebibliography}{99}

\bibitem{cit1} a) J. A. Wilson, F. J. DiSalvo, and F. Mahajan, Phys.
Rev.  Lett.  \textbf{32}, 882 (1974); Adv.  Phys.  \textbf{24}, 117
(1975).  b) D. E. Moncton, J. D. Axe, and F. J. DiSalvo, Phys.  Rev.
B\textbf{ 16}, 801(1977).

\bibitem{cit2} T. M. Rice and G. K. Scott, Phys.  Rev.  Lett.
\textbf{35}, 120 (1975).

\bibitem{cit5} A. S. Barker, Jr., J. A. Ditzenberger, and F. J.
DiSalvo, Phys.  Rev.  B \textbf{12}, 2049 (1975).

\bibitem{cit6} R. V. Coleman, B. Giambattista, P. K. Hansma, A.
Johnson, W. W. McNairy and C. G. Slough, Adv.  Phys.  \textbf{37}, 559
(1988).

\bibitem{cit7} W. L. McMillan, Phys.  Rev.  B \textbf{16}, 643 (1977).

\bibitem{cit3} M. D. Johannes and I. I. Mazin, Phys.  Rev.  B
\textbf{77}, 165135 (2008).

\bibitem{cit18} A. B. Migdal, Zh.  Eksp.  Teor.  Fiz.  \textbf{34},
1438 (1958) [Sov.  Phys.  JETP \textbf{37}, 996 (1958)].

\bibitem{cit4} C. M. Varma and A. L. Simons, Phys.  Rev.  Lett.
\textbf{51}, 138 (1983).

\bibitem{cit19} S. V. Borisenko, A. A. Kordyuk, V. B. Zabolotnyy, D. S.
Inosov, D. Evtushinsky, B. Buechner, A. N. Yaresko, A. Varykhalov, R.
Follath, W. Eberhardt, L. Patthey, and H. Berger, Phys.  Rev.  Lett.
\textbf{102}, 166402 (2009).

\bibitem{cit24} R. Liu, C. G. Olson, W. C. Tonjes, and R. F. Frindt,
Phys.  Rev.  Lett.  \textbf{80}, 5762 (1998); R. Liu et al., Phys.
Rev.  B \textbf{61}, 5212 (2000).

\bibitem{cit9} a) Clare C. Yu and P. W. Anderson, Phys.  Rev.  B
\textbf{29,} 6165 (1984); b) in \emph{Proceedings of the International
School of Physics ``Enrico Fermi''}, eds.  F. Bassani, F. Fumi, and M.
Tossi, 1983.

\bibitem{cit8} W. L. McMillan, Phys.  Rev.  B \textbf{12}, 1187
(1975).

\bibitem{cit10} M. Marezio, P. D.Dernier, A. Menth, and G.W. Hull,
Jr., J. Solid State Chem.\textbf{4}, 425 (1972).

\bibitem{cit11} F. Weber, S. Rosenkranz, J.-P. Castellan, R. Osborn,
R. Hott, R. Heid, K.-P. Bohnen, T. Egami, A. H. Said, D. Reznik, Phys.
Rev.  Lett., \textbf{107}, 107403 (2011).

\bibitem{cit12} A. A. Abrikosov, L. P. Gorkov and I. E. Dzyaloshinski,
\textit{Methods of Quantum Field Theory in Statistical Physics},
Dover, New York, 1963.

\bibitem{cit13} T. Holstein, \textit{Ann.  Phys.} \textbf{8}, 343
(1959).

\bibitem{cit14} R. Corcoran, P. Meeson, Y. Onuki, P.-A. Probst, M.
Springford, K. Takita, H. Harima, G. Y. Guo, and B. L. Gyorffy, J.
Phys.: Condens.  Matter \textbf{6}, 4479 (1994).

\bibitem{cit15} Y. Onodera, Prog. Theor. Phys.\textbf{44}, 1477 (1970)

\bibitem{cit25} D. W. Shen, Y. Zhang, L. X. Yang, J. Wei, H. W. Ou, J. K.
Dong, B. P. Xie, C. He, J. F. Zhao, B. Zhou, M. Arita, K. Shimada, H.
Namatame, M. Taniguchi, J. Shi, and D. L. Feng, Phys.  Rev.  Lett.
\textbf{101}, 226406 (2008); the authors, however, claimed that ``the
states in these low-DOS regions … play a primary role in the CDW
formation".

\bibitem{cit16} V. G. Vaks, V. M. Galitskii, and A. I. Larkin, Zh.
Eksp. Teor.  Fiz.\textbf{51}, 1592 (1966) [Sov.  Phys.  - JETP
\textbf{24}, 1071 (1967)].

\bibitem{cit22} a) R. A. Craven and S. F. Meyer, Phys.  Rev.  B
\textbf{8}, 4583 (1977); b) A. LeBlanc and A. Nader, Solid State
Commun.  \textbf{150}, 1346 (2010).

\bibitem{cit23} T. Valla, A. V. Fedorov, P. D. Johnson, J. Xue, K. E.
Smith, and F. J. DiSalvo, Phys.  Rev.  Lett.  \textbf{85}, 4759
(2000).

\bibitem{cit26} T. Valla, A. V. Fedorov, and P. D. Johnson, P.-A.
Glans, C. McGuinness, K. E. Smith, E. Y. Andrei, and H. Berger, Phys.
Rev.  Lett.  \textbf{92}, 086401 (2004).

\bibitem{cit20} A. Georges, G. Kotliar, W. Krauth, and M. J.
Rozenberg, Rev.  Mod.  Phys.  \textbf{68}, 13 (1996).

\bibitem{cit21} D. Meyer, A. C. Hewson, and R. Bulla, Phys.  Rev.
Lett.  \textbf{89}, 196401 (2002).

\end{thebibliography}
\end{document}